# Magnetostructural martensitic transformations with large volume changes and magneto-strains in all-*d*-metal Heusler alloys

Z. Y. Wei,[1] E. K. Liu,[1,a)] Y. Li,[1,2] X. L. Han,[3] Z. W. Du,[3] H. Z. Luo,[2] G. D. Liu,[2] X. K. Xi,[1] H. W. Zhang,[1] W. H. Wang,[1] G. H. Wu[1]

[1] *State Key Laboratory for Magnetism, Beijing National Laboratory for Condensed Matter Physics, Institute of Physics, Chinese Academy of Sciences, Beijing 100190, China*

[2] *School of Materials Science and Engineering, Hebei University of Technology, Tianjin 300130, China*

[3] *National Center of Analysis and Testing for Nonferrous Metals and Electronic Materials, General Research Institute for Nonferrous Metals, Beijing 100088, China*

Abstract: The all-*d*-metal $Mn_2$-based Heusler ferromagnetic shape memory alloys $Mn_{50}Ni_{40-x}Co_xTi_{10}$ ($x$ = 8 and 9.5) are realized. With a generic comparison between *d*-metal Ti and main-group elements in lowering the transformation temperature, the magnetostructural martensitic transformations are established by further introducing Co to produce local ferromagnetic Mn-Co-Mn configurations. A 5-fold modulation and (3, -2) stacking of [00 10] of martensite are determined by XRD and HRTEM analysis. Based on the transformation, a large magneto-strain of 6900 ppm and a large volume change of -2.54% are observed in polycrystalline samples, which makes the all-*d*-metal magnetic martensitic alloys of interest for magnetic/pressure multi-field driven applications.

[a)] E-mail: ekliu@iphy.ac.cn

Heusler alloy is a large family of materials that exhibit diverse functionalities including half-metallicity,[1] Hall effect,[2] magnetoresistance,[3] shape memory effect,[4] magnetocaloric effect (MCE),[5] and energy conversion.[6] The main-group elements are supposed to be necessary in the composition of Heusler alloys because the *p*-*d* hybridization between main-group atoms and their nearest-neighbor transition-metal atoms is of great importance for the formation of the Heusler phase structure.[7,8] Recently, all-*d*-metal Heusler alloys $Ni_{50}Mn_{50-y}Ti_y$ and $Mn_{50}Ni_{50-y}Ti_y$ were reported from our group.[9] The *d*-*d* hybridization from transition-metal elements carrying low valence-electrons (e.g., Ti) are proved to act as the similar role as *p*-*d* hybridization in forming Heusler phase. In addition, the all-*d*-metal Heusler alloys $Ni_{50-x}Co_xMn_{50-y}Ti_y$ were developed as a kind of ferromagnetic shape memory alloys (FSMAs) with martensitic transformations (MTs), which are highly desired in wide research area as smart materials including actuating, magnetic cooling, magnetostriction, and energy conversion. Among the multi-functionalities of FSMAs, the large magnetic-field-induced strain is one of the important pursued properties. Large magnetic-field-induced strains have been reported mostly in single-crystal Heusler alloys, such as NiMnGa (6%),[10] NiMnGa-based (12%),[11] NiCoMnIn (3%),[4] NiCoMnSn (1%)[12]. The large strain is an outcome of lattice expansion along specific axis or contraction along the others when MT takes place. The first-order MT is always accompanied by a volume discontinuity ($\Delta V$). Large $\Delta V$ provides another degrees of freedom for external stimulin, which can bring mechanocaloric effects such as barocaloric in hydrostatic pressure and elastocaloric in uniaxial stress.[13,14]

In this study, we constructed magnetostructural martensitic transformations around room temperature in all-*d*-metal $Mn_{50}Ni_{40-x}Co_xTi_{10}$ Heusler alloys. The martensite with a 5-layer modulated structure with a (3, -2) stacking of [00 10]5M is determined. The volume change during the MT is found as large as -2.54%. A large magnetic field-induced strain of 6900 ppm is obtained in polycrystalline samples.

$Mn_{50}Ni_{40-x}Co_xTi_{10}$ ($x$ = 8.0, 9.5; denoted as Co$x$) alloys were prepared by arc melting high purity metals in argon atmosphere. The ingots were afterwards annealed at 1103 K in evacuated quartz tubes for six days and then quenched in cold water.

Room temperature (RT) X-ray diffraction (XRD) was performed using Cu-$K\alpha$ radiation. The high-resolution (HR) images and selected area electron diffractions (SAED) of martensite variants were performed on a transmission electron microscope (TEM). Magnetic properties were measured in a superconductive quantum interference device (SQUID) magnetometer. The martensitic and magnetic transition temperatures were determined by differential scanning calorimetry (DSC) and magnetic measurements. The strains were measured by a strain gauge on a physical property measurement system (PPMS).

The *d*-metal Ti has been proved to show similar effect to main-group (*p*-block) elements on forming and stabilizing the all-*d*-metal B2 Heusler structure, where *d*-*d* hybridizations between Ti at D site and its nearest-neighbor *d*-metal atoms at A(C) site are formed.[9] Figure 1 shows the valence electron concentration (*e/a*) dependence of martensitic transformation temperature ($T_t$) for all-*d*-metal $Ni_{50}Mn_{50-y}Ti_y$ ($0 < y \leq 20$), $Mn_{50}Ni_{50-y}Ti_y$ ($0 < y \leq 15$) and NiMn-based conventional Heusler alloys. The various slopes indicate different abilities of elements at D site to stabilize the parent phase, that is, the abilities to tailor high-temperature MTs to lower temperatures. In the case of $Ni_{50}Mn_{50-y}X_y$, the slope for Ti is higher than that for Al and Ga, close to In, and lower than Sn and Sb. This suggests that *d*-metal Ti with *d*-*d* hybridization as well as its size factor, can produce similar behavior in stabilizing the parent phase in $Mn_{50}Ni_{50-y}X_y$ systems, compared with the main-group elements with *p*-*d* hybridizations in conventional Heusler alloys.[19-21] However, both $Ni_{50}Mn_{50-y}Ti_y$ and $Mn_{50}Ni_{50-y}Ti_y$ systems are paramagnetic (PM) state when MTs take place and their martensites also show weak magnetizations.

In order to establish magnetostructural martensitic transformations, we further applied the "FM activation effect" of Co atom[9,22-28] to the $Mn_{50}Ni_{50-y}Ti_y$ ($0 < y \leq 15$) system. Based on the valence-electron site occupation rule in Heusler alloys,[7,28] Ni-rich $Mn_{50}Ni_{50-y}Ti_y$ in parent state has an off-stoichiometric $Hg_2CuTi$-type structure with 4 face-center-cubic sublattices (denoted as A, B, C and D along the body diagonal line), as shown in Fig. 2a. The atom occupation form should be written as $(Mn_yNi_{25-y})_A(Mn_{25})_B(Ni_{25})_C(Ti_yMn_{25-y})_D$ ($0 < y \leq 15$). The additional $Ni_{25-y}$ atoms take

MnA site and drive commensurate $Mn_{25-y}$ atoms to TiD site, leading to antiferromagnetic (AFM) coupling between MnB and MnD.[29,30] According to the requirement of the atom configuration for "FM activation effect", Co should be introduced to chemically substitute Ni atoms at A or C site to form the local MnB-Co(A/C)-MnD configurations, which have been proved to exhibit FM coupling with parallel magnetic moments.[28,29] We performed the theoretical calculations using the first-principles total-energy method (see Supplementary material[18]). The results reveal that Co prefers to replace Ni at A site, showing a clear trend of separate occupation of Co and Ni atoms at A and C sites. Therefore, as shown in Fig. 2a, the ideal atom occupation form of the studied $Mn_{50}Ni_{40-x}Co_xTi_{10}$ ($y$ = 10, $x$ = 8 and 9.5) is $(Mn_{10}Ni_{15-x}Co_x)_A(Mn_{25})_B(Ni_{25})_C(Ti_{10}Mn_{15})_D$ and the MnB-CoA-MnD configuration is formed in the alloys. The calculations further assemble the parallel moments for MnB-CoA-MnD configuration (MnB ~2.83 $\mu_B$, CoA ~1.29 $\mu_B$ and MnD ~2.99 $\mu_B$ for Co9.5 alloy, see Table S1 in Supplementary material[18]).

Figure 2(b) presents the thermo-magnetization (M-T) curves of $Mn_{50}Ni_{40-x}Co_xTi_{10}$ ($y$ = 10, $x$ = 8 and 9.5) in a magnetic field of 100 Oe. One can see the Curie temperature ($T_C^A$) of parent phase appears above $T_t$ and increases with increasing Co content. In a field of 50 kOe, a magnetization about 85 emu/g for Co9.5 alloy can be observed before the MT. The desired strong ferromagnetism and magnetostructural transformation are thus established in $Mn_{50}Ni_{40-x}Co_xTi_{10}$ system, with the aid of the "FM activation effect" of Co. For the Co8 alloy, $T_t$ (330 K on cooling) is so close to $T_C^A$ (360 K) that the FM coupling in parent phase cannot be built sufficiently when MT occurs. In contrast, $T_t$ (298 K on cooling) and $T_C^A$ (400 K) are well separated for Co9.5 alloy with higher Co content. Co doping not only promotes the FM coupling in parent phase but lowers $T_t$, similar to the cases in many Heusler alloys.[23-25,27] Both MTs of two samples show a change from FM parent phase to weak-magnetism martensite, with a significant magnetization difference ($\Delta$M). In a magnetic field of 50 kOe, Co9.5 alloy shows a large $\Delta$M of 80 emu/g across the MT (Fig. 2(b)). This large $\Delta$M would stabilize the FM parent phase by offering more Zeeman energy[31] and would result in a considerable shift of $T_t$ according to Clausius-Clapeyron relation $dT/dH = -\mu_0\Delta M/\Delta S$,[4] as observed experimentally with $\Delta T$ = 28 K for 50-kOe field.

RT XRD patterns of Co8 and Co9.5 polycrystalline alloys are shown in Fig. 3a. Two sets of patterns are observed for Co9.5 alloy. One is cubic B2 parent phase while the other is identified as 5-fold modulated (5M) structure martensite. The widening of diffraction peaks results from the largely distorted lattice planes upon MT.[32] A small amount of non-modulated $L1_0$ phase is seen in the XRD pattern of Co8, which is of the same structure as binary martensitic NiMn alloy.[33] This coexistence of modulated and non-modulated structures can be also observed in Ni-Mn-Ga alloys.[34-37] In both alloys, the coexistence of parent and martensite phases implies the MTs occur around room temperature, which is in agreement with the *M-T* curves. The higher proportion of B2 phase for Co9.5 compared to Co8 seen from XRD pattern is due to the lower $T_t$ of Co9.5 compared to Co8. Based on the coexistence of parent and 5M structures, the lattice parameters of Co9.5 are refined as $a_{cubic}$ = 5.916 Å, $a_{5M}$ = 4.364 Å, $b_{5M}$ = 5.437 Å, $c_{5M}$ = 21.296 Å, and $\beta_{5M}$ = 93.24°, and Co8, $a_{cubic}$ = 5.915 Å, $a_{5M}$ = 4.376 Å, $b_{5M}$ = 5.420 Å, $c_{5M}$ = 21.295 Å, $\beta_{5M}$ = 93.26°, and $a_{L10}$ = 5.218 Å, $c_{L10}$ = 7.380 Å.

Figure 3(b) shows the HR-TEM images of Co8 alloy. The periodic martensite variants reflected by black and white contrasts are observed. Enlarged image in the inset of Fig. 3(b) shows the periodic structure has a thickness of five atomic planes in one period and is composed of (3, -2) stacking of $[0010]_{5M}$ plane. The corresponding SAED presents four satellites between the main reflections (the direction of the electron beam is along the $[-111]_{cubic}//[210]_{5M}$), which confirms the 5M structure of martensite, and is consistent with the XRD analysis. A little fraction of six or other numbers are also observed, which are probably from other modulated structures.[37] This irregularity would account for the elongated satellites in the SAED pattern. Moreover, applying the lattice parameters obtained from XRD analysis to the simulation of the electron diffraction patterns of 5M martensite, the consistent result is obtained with the observed SAED patterns in inset of Fig. 3(b).

Figure 4(a) shows the temperature dependence of strain for polycrystalline Co9.5 alloy in applied magnetic fields of 0 and 120 kOe. The profiles of the curves are similar to the $M$-$T$ curve in Fig. 2(b). $T_t$ and reverse $T_t$ are 303 K and 318 K, respectively, which cohere with those obtained from $M$-$T$ curves within the measurement accuracy. A field of 120 kOe can drive the MT to lower temperatures with ΔT = 29 K. Remarkably, a rather large strain up to 8150 ppm across the MT is obtained. To eliminate the strain enhanced by possible grain orientation, we prepared another sample and measured strains along two directions that are perpendicular to each other. The results show the strain difference along the two directions is less than 300 ppm (< 4%), which indicates that the large strain is almost isotropic. This value is a quite large one for a polycrystalline FSMA. Based on the MT, a large magneto-strain of 6900 ppm under a field change of 120 kOe is observed at 305 K, as shown in Fig. 4(b).

The large strains shown in Fig. 4 originate from the distortion of the cubic parent phase during the MT. The unit cell will elongate along specific crystallographic orientations and shorten along the other ones. For the present system, $[-110]_{cubic}/[100]_{5M}$ is the elongated direction during the transformation to 5M martensite. For an isotropic polycrystalline sample the lattice strain $\varepsilon$ resulted from MT can be calculated from the volume change by the formula,

$$\varepsilon = \frac{\Delta L}{L} = 1 - \sqrt[3]{\frac{V_{5M}}{2.5V_{cubic}}}$$

where $V_{5M}$ and $V_{cubic}$ are the unit-cell volumes of 5M structure and cubic parent phase respectively. The coefficient 2.5 is the ratio between crystal cell volumes of two phases. Using the lattice parameters of parent phase and 5M martensite of Co9.5 alloy, $V_{5M} = abc\sin\beta$ = 504.484 Å$^3$, $V_{cubic}$ = 206.949 Å$^3$, the volume shrink is $\Delta V/V = (V_{5M} - 2.5V_{cubic})/(2.5V_{cubic})$ = -2.54% and the consequent strain is $\varepsilon$ = 8540 ppm, which is quite close to the measured value of $\varepsilon$ = 8150 ppm. The consistent results by different methods further indicate the polycrystalline Co9.5 specimen has almost isotropic microstructure. The large strain comes from the crystal volume shrink rather than the striction along a specific crystal orientation. The simple fabrication of the polycrystalline specimens would make the materials convenient for practical applications.

Figure 5 shows the volume changes and corresponding maximum strains upon MTs for some Heusler and other FSMAs. The all-*d*-metal Co9.5 alloy exhibits both very large volume discontinuity and large lattice parameter discontinuity with -2.54% and -8.1% respectively, compared to conventional Heusler alloys that undergo MTs to modulated martensite, and to other SMAs like FePt and NiTi. It should be mentioned that, after MT the studied samples remain a bulk and no crack is formed due to high mechanical toughness originated from d-d covalent bonding in these all-d-metal alloys. For Co8 alloy containing the non-modulated $L1_0$ martensite the maximal lattice parameter discontinuity reaches up to 24.8%. We notice that the volume change of Co8 alloy upon the MT to a modulated martensite is lower than that to $L1_0$ structure (Tables S2 in Supplementary material[18]). This can be explained by the concept of adaptive modulation[34,39] which considers the modulated martensite as specific arrangements of nano-twin lamellae of simple non-modulated martensite. In other words, the volume of modulated martensite should be close to that of non-modulated one while the maximal strain for modulated martensite is greatly reduced by adaptive martensite, in order to decrease the stress upon MT. According to Clausius-Clapeyron relation $dT/dp = \Delta V/\Delta S$, a volume/lattice-constant discontinuity will lead to pressure/stress-driven MT.[31] In SMAs, both barocaloric and elastocaloric effects can be obtained,[13,40] which are especially expected in our all-*d*-metal Heusler FSMAs with a large volume/lattice-constant discontinuity. Meanwhile, the FMMTs of these magnetic materials can be induced by a magnetic field, which usually produces a giant MCE. These characters jointly widen the applications of these materials such as multi-caloric cooling.[5,14,41,42]

To conclude, the low valence-electron *d*-metal Ti element shows consistent effects with the main-group elements in forming and stabilizing Heusler phase. The coherent decreasing trend in martensitic transformation temperature between Ti and In elements is observed. In the developed all-*d*-metal Heusler FSMA $Mn_{50}Ni_{40-x}Co_xTi_{10}$, the magnetostructural martensitic transformation, producing a 5-fold modulated martensite, exhibits large magneto-strain of 6900 ppm and large volume discontinuity of -2.54%, which may efficiently expand the degrees of freedom of stimulating field to pressure/stress from the temperature and magnetic field.

This work was supported by National Natural Science Foundation of China (51301195, 51431009, 51271038 and 51471184), Beijing Municipal Science and Technology Commission (Z141100004214004), and Youth Innovation Promotion Association of Chinese Academy of Sciences (2013002).

Captions of Figures

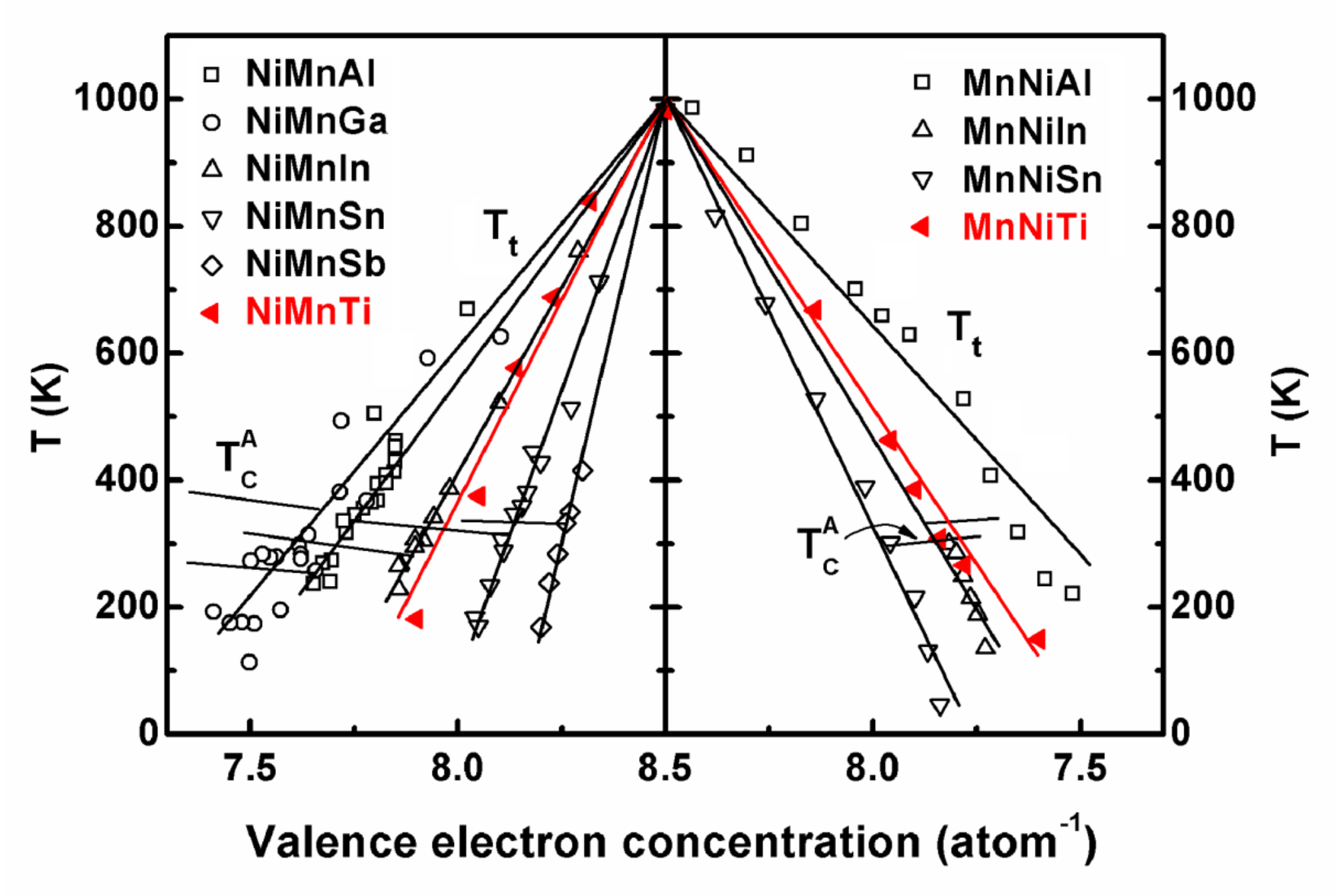

FIG. 1. Valence electron concentration *e/a* dependence of martensitic transformation temperature ($T_t$) for some typical Heusler alloys: $Ni_{50}Mn_{50-y}X_y$ (left) (Some data are taken from refs. 15, 16 and 17) and $Mn_{50}Ni_{50-y}X_y$ (right) system. Here, X = Al, Ga, In, Sn, Sb, and Ti. (More references can be found in Supplementary material.[18])

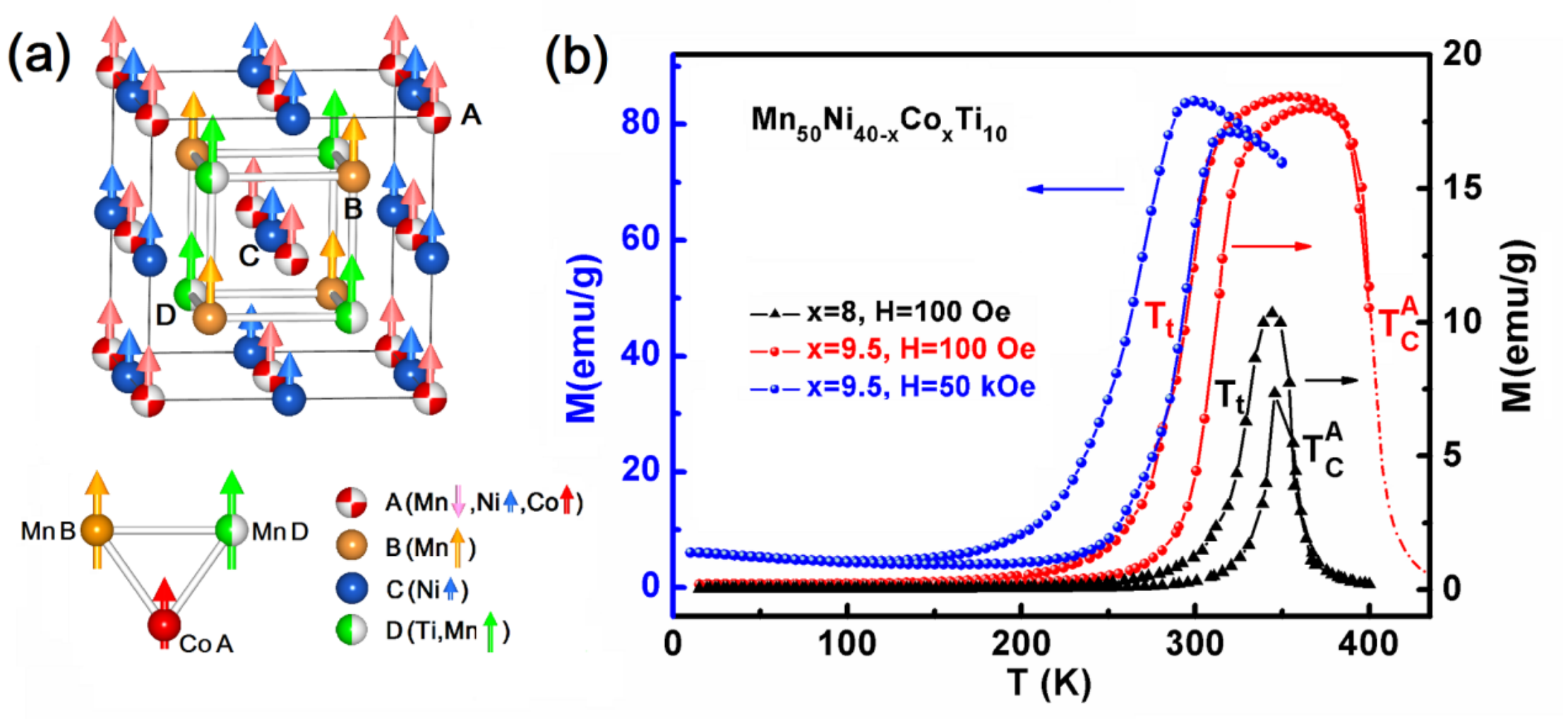

FIG. 2. (a) Schematic structure of $Mn_{50}Ni_{40-x}Co_xTi_{10}$ and MnB-CoA-MnD local configurations in the alloys. (b) *M*-*T* curves of $Mn_{50}Ni_{40-x}Co_xTi_{10}$ (Co*x*, *x* = 8.0 and 9.5) samples. *M*-*T* curves in a magnetic field of 100 Oe for Co8 and Co9.5 alloys are shown. *M*-*T* curve in magnetic field of 50 kOe for Co9.5 alloy is shown.

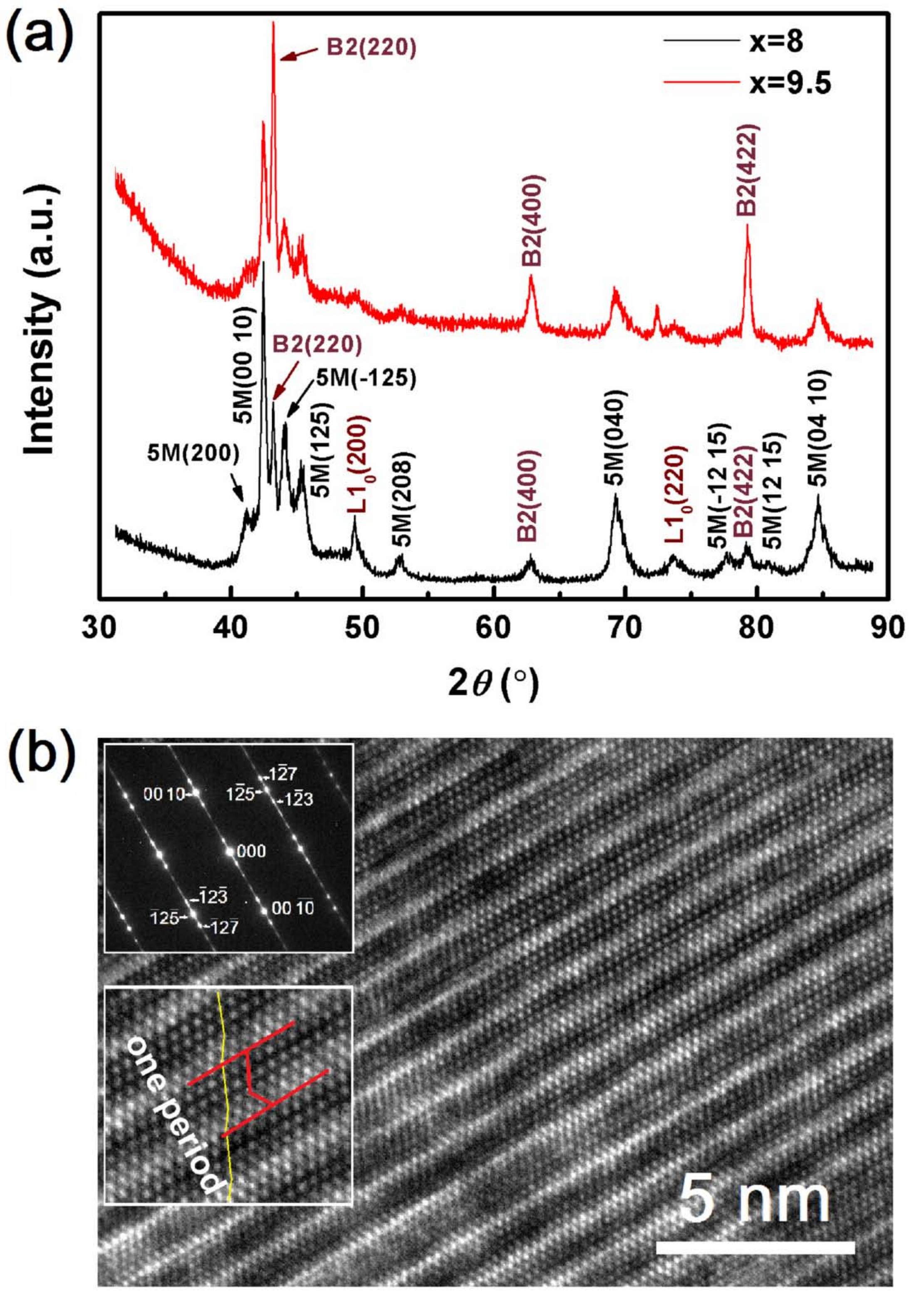

FIG. 3. (a) RT XRD patterns of $Mn_{50}Ni_{32}Co_{8}Ti_{10}$ and $Mn_{50}Ni_{30.5}Co_{9.5}Ti_{10}$ polycrystalline samples. (b) HRTEM image of $Mn_{50}Ni_{32}Co_{8}Ti_{10}$ presenting the modulated martensite structure. Insets are enlarged image and corresponding SAED image. The yellow line is a guide for eyes.

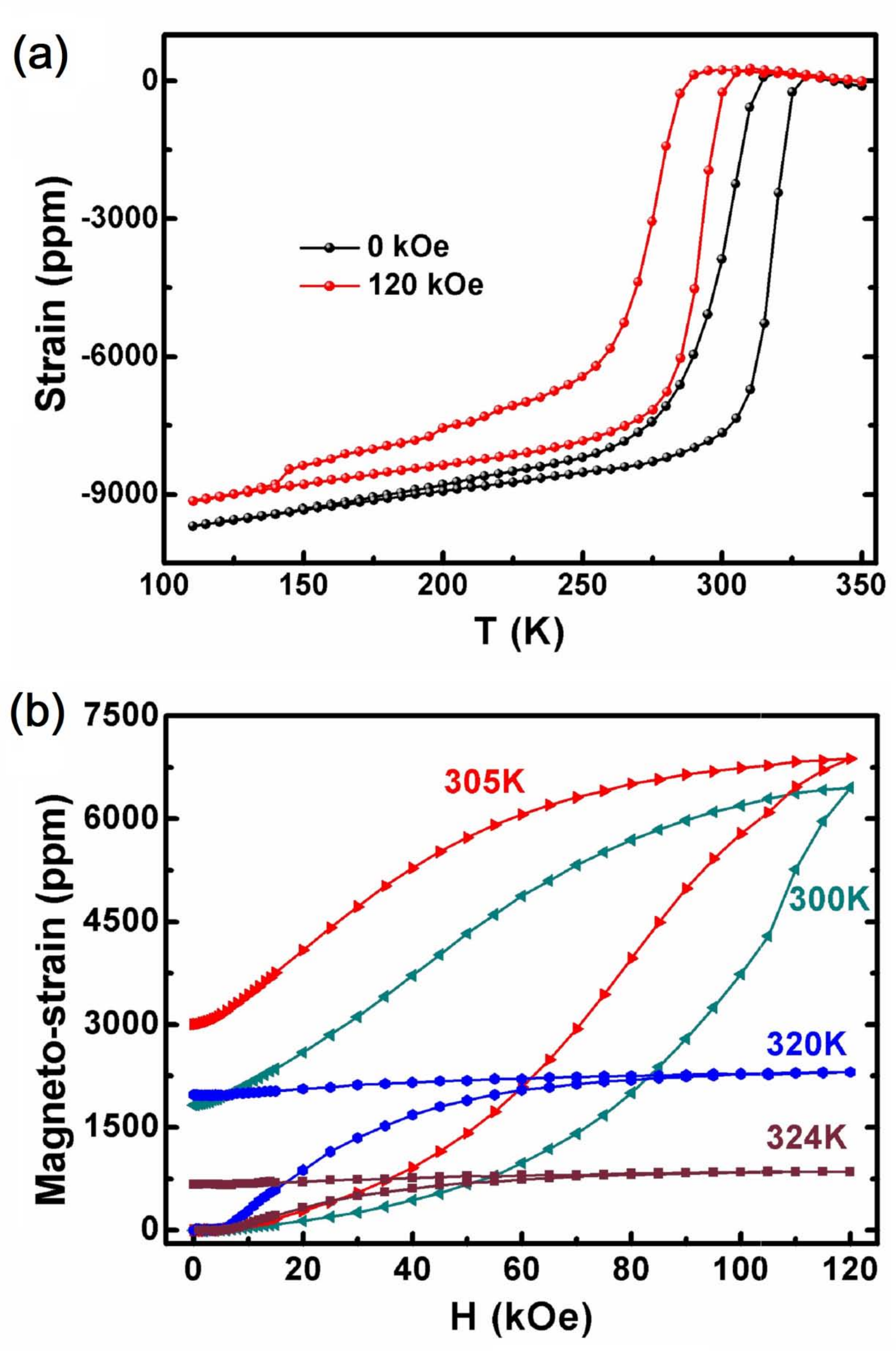

FIG. 4. (a) Temperature dependence of strain for polycrystalline Co9.5 in magnetic fields of 0 and 120 kOe. (b) Magneto-strain at temperatures around MT of Co9.5 sample.

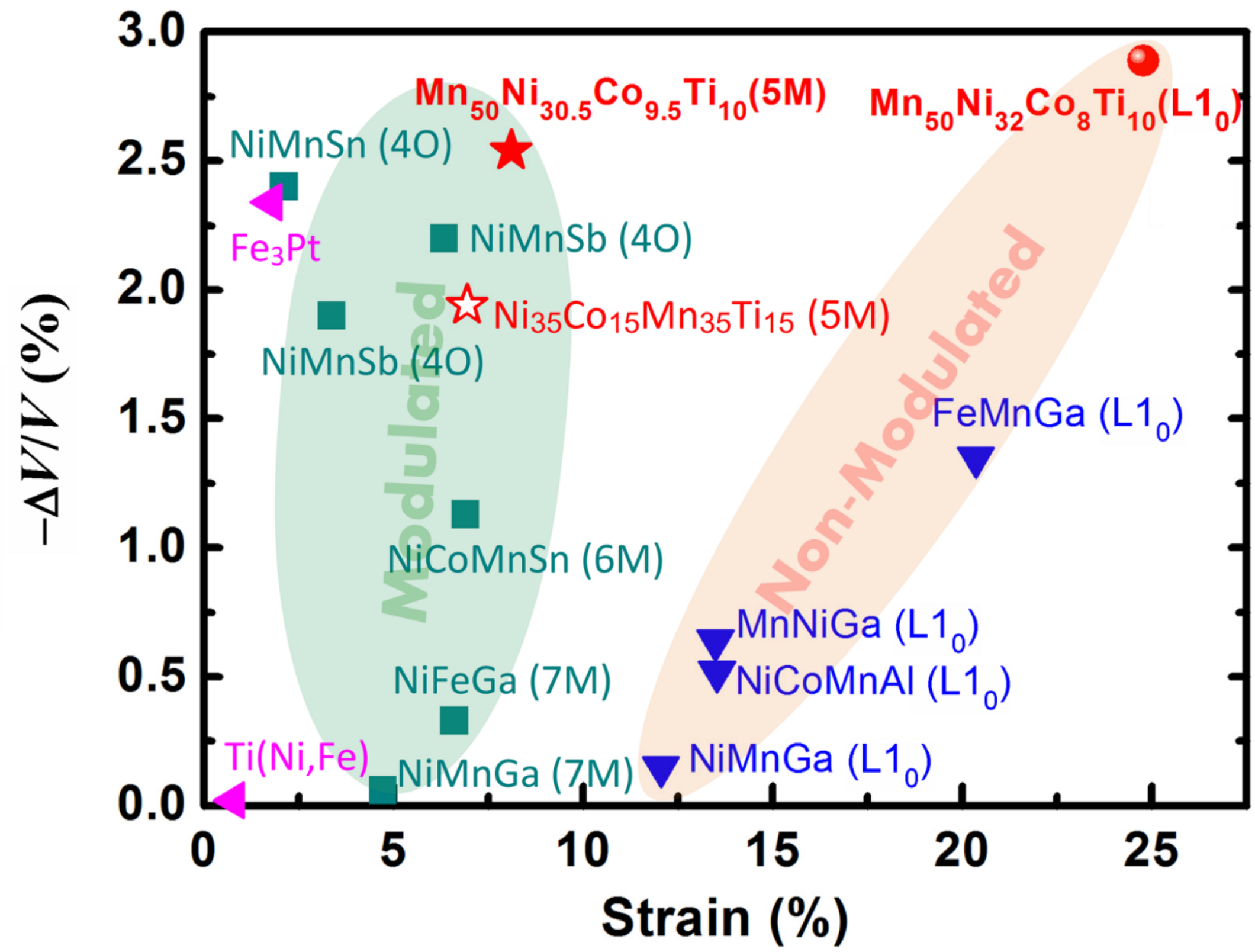

FIG. 5. Volume changes ($\Delta V/V$) and corresponding maximum strains upon martensitic transformations for some Heusler and other FSMAs (★, ☆ and ● denote the all-$d$-metal FSMAs). The 4O modulated structure is a special case of incommensurate 5M modulated structure as modulation vector $q$ approaches 0.5.[38] (Detailed data and corresponding references can be found in Tables S2 and S3 in Supplementary material.[18])

# Supplementary material to

## Magnetostructural martensitic transformations with large volume changes and magneto-strains in all-*d*-metal Heusler alloys

Z. Y. Wei,[1] E. K. Liu,[1,a)] Y. Li,[1,2] X. L. Han,[3] Z. W. Du,[3] H. Z. Luo,[2] G. D. Liu,[2] X. K. Xi,[1] H. W. Zhang,[1] W. H. Wang,[1] G. H. Wu[1]

[1] *State Key Laboratory for Magnetism, Beijing National Laboratory for Condensed Matter Physics, Institute of Physics, Chinese Academy of Sciences, Beijing 100190, China*

[2] *School of Materials Science and Engineering, Hebei University of Technology, Tianjin 300130, China*

[3] *National Center of Analysis and Testing for Nonferrous Metals and Electronic Materials, General Research Institute for Nonferrous Metals, Beijing 100088, China*

**Detail information in Figure 1.** The data in Figure 1 are taken from literature, NiMnAl,[1-3] NiMnGa,[4] NiMnIn,[5] NiMnSn,[5-7], NiMnSb,[5] MnNiAl,[8] MnNiIn,[9] and MnNiSn[10].

**First-principles calculations.** For the off-stoichiometric $Hg_2CuTi$-type $Mn_{50}Ni_{40}Ti_{10}$ the atom occupation is $(Mn_{10}Ni_{15})_A(Mn_{25})_B(Ni_{25})_C(Ti_{10}Mn_{15})_D$ according to valence-electron site occupation rule. When Ni atoms are substituted by Co atoms, Co atoms may occupy either A site or C site or randomly. In order to investigate the occupation of the substituting Co for Ni, first-principles calculations were performed on $Mn_{50}Ni_{30.5}Co_{9.5}Ti_{10}$ by using the full-potential Korringa-Kohn-Rostoker (KKR) Green's function method combined with coherent potential approximation (CPA). [11-13] The exchange correlation is used as mjw. Considering the valence electron occupation rule, the formula of the alloy can be written as $(Mn_{10}Ni_{5.5+z}Co_{9.5-z})_A(Mn_{25})_B(Ni_{25-z}Co_z)_C(Ti_{10}Mn_{15})_D$ when Co is introduced. By changing Co content ($z$) at C site, we could obtain the possible minimum of energy in a series of total energies, to search the most stable state with a certain $z$ value. For each $z$ value, the equilibrium lattice constant, with a lowest energy, was obtained by geometrical optimization, which is shown in energy - lattice curve of Figure S1.

a) E-mail: ekliu@iphy.ac.cn

TABLE S1 and Figure S2 show the calculated total energy, equilibrium lattice constant and total formula (atom) magnetic moments as functions of Co content ($z$) on C site. One can see that the total energy of the system increases with increasing Co content ($z$), which indicates that the introduced Co atoms prefer the energetically stable A site, rather than C site. The calculated results indicate that there exists a clear trend of separate occupation of Co and Ni atoms at A and C sites. For $(Mn_{10}Ni_{15})_A(Mn_{25})_B(Ni_{25})_C(Ti_{10}Mn_{15})_D$, the introduced Co atoms prefer to replace the Ni atoms at A site, leaving Ni atoms at C site.

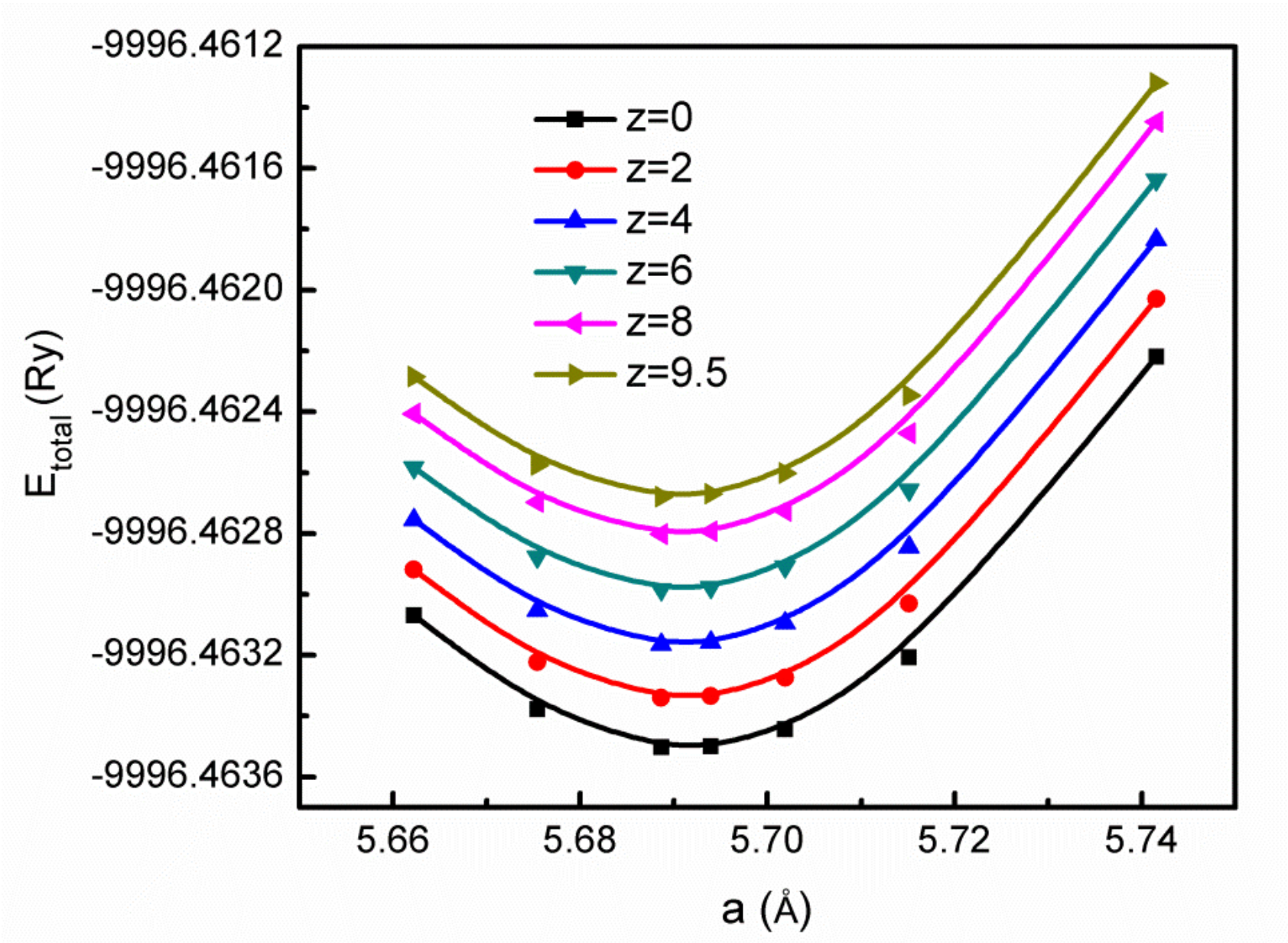

Figure S1. Lattice parameter dependence of calculated total energy for various Co content (z) in $(Mn_{10}Ni_{5.5+z}Co_{9.5-z})_A(Mn_{25})_B(Ni_{25-z}Co_z)_C(Ti_{10}Mn_{15})_D$.

**TABLE S1. Calculated results of lattice constant ($a$, in Å), total energy ($E_{total}$, in Ry), and magnetic moments ($M_{total}$, m, in $\mu_B$) of $(Mn_{10}Ni_{5.5+z}Co_{9.5-z})_A(Mn_{25})_B(Ni_{25-z}Co_z)_C(Ti_{10}Mn_{15})_D$.**

| $z$ | $a$ | $E_{total}$ | $M_{total}$ | $m_{MnA}$ | $m_{NiA}$ | $m_{CoA}$ | $m_{MnB}$ | $m_{NiC}$ | $m_{CoC}$ | $m_{TiD}$ | $m_{MnD}$ |
|---|---|---|---|---|---|---|---|---|---|---|---|
| 0 | 5.689 | -9996.46350 | 5.4209 | -0.8415 | 0.5912 | 1.29287 | 2.8322 | 0.5059 | -- | -0.0446 | 2.9860 |
| 2 | 5.689 | -9996.46334 | 5.4330 | -0.8172 | 0.5906 | 1.2902 | 2.8361 | 0.5070 | 1.1391 | -0.0448 | 2.9873 |
| 4 | 5.689 | -9996.46316 | 5.4424 | -0.7970 | 0.5895 | 1.2870 | 2.8395 | 0.5078 | 1.1415 | -0.0450 | 2.9884 |
| 6 | 5.689 | -9996.46298 | 5.4497 | -0.7803 | 0.5882 | 1.2835 | 2.8426 | 0.5085 | 1.1442 | -0.0450 | 2.9894 |
| 8 | 5.689 | -9996.46280 | 5.4559 | -0.7664 | 0.5870 | 1.2797 | 2.8456 | 0.5090 | 1.1472 | -0.0448 | 2.9903 |
| 9.5 | 5.689 | -9996.46268 | 5.4603 | -0.7573 | 0.5862 | -- | 2.8479 | 0.5094 | 1.1494 | -0.0445 | 2.9910 |

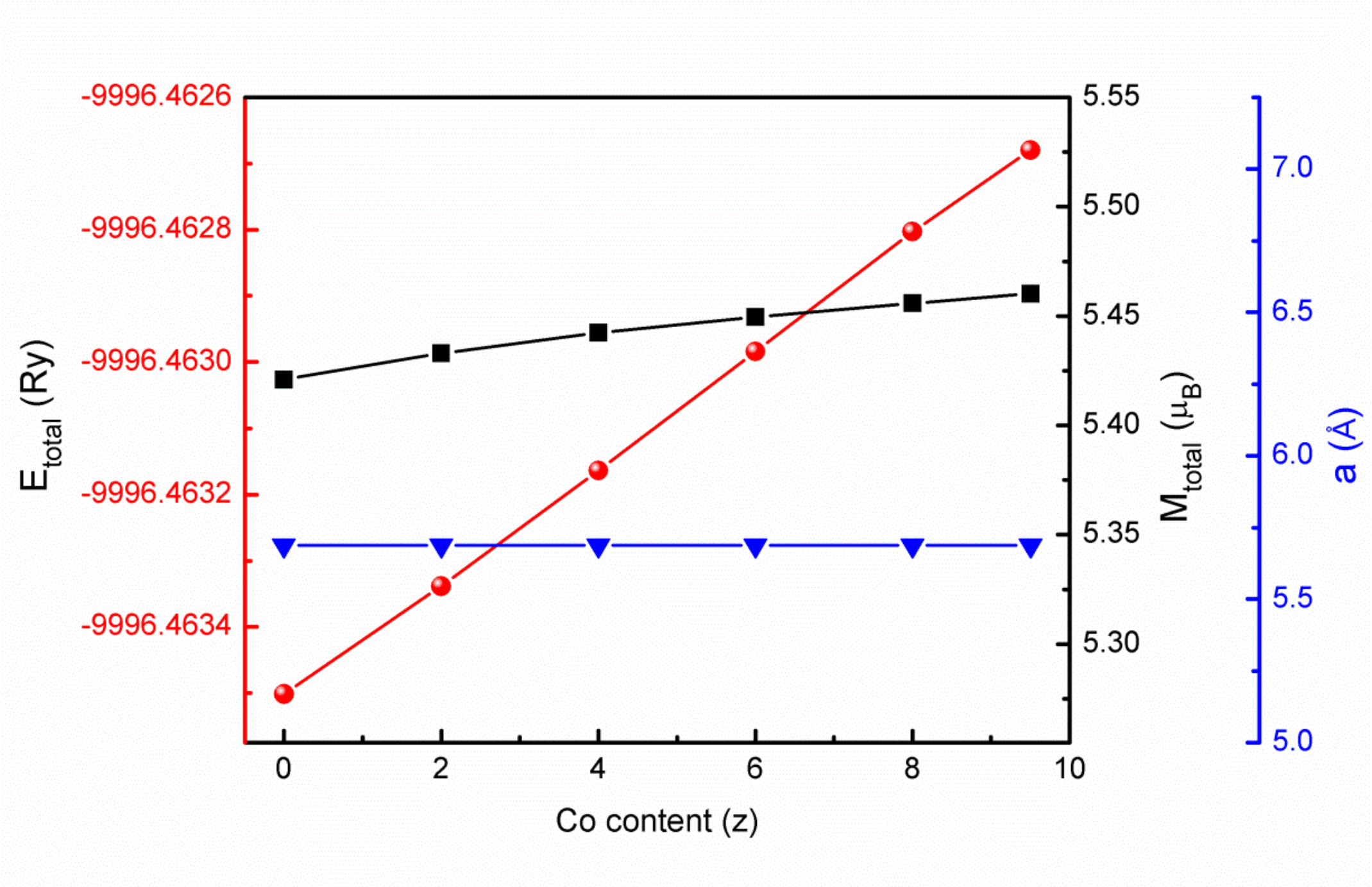

Figure S2. The calculated total energy and corresponding lattice constant and total formula magnetic moment as functions of substituting Co content ($z$) for $(Mn_{10}Ni_{5.5+z}Co_{9.5-z})_A(Mn_{25})_B(Ni_{25-z}Co_z)_C(Ti_{10}Mn_{15})_D$.

**Summary of lattice parameters, uniaxial strains and volume changes** of Heusler alloys and some typical shape memory alloys, as listed in TABLEs S2 and S3, respectively. The strains are calculated based on corresponding lattice parameters of martensite to that of parent phase. $\varepsilon$ refers to maximum strain based on the difference of two axial in martensite.

**TABLE S2. Lattice constants, uniaxial strains, volume changes ($\Delta V/V$) of all-*d*-metal Heusler alloys, conventional Heusler alloys.**

| Alloys | Parent | Martensite | | | | | | | |
|---|---|---|---|---|---|---|---|---|---|
| | $a_0$ | *a* (strain) | *b* (strain) | *c*/5 (strain) | $\beta$ | Sym. | \|ε\| (%) | $\Delta V/V$ (%) | Ref. |
| | | $0.5\sqrt{2}a_0$ | $a_0$ | $0.5\sqrt{2}a_0$ | | | | - | |
| $Ni_{36.5}Co_{13.5}Mn_{35}Ti_{15}$ | 5.8952 | 4.320 (3.64%) | 5.486 (-6.94%) | 4.2436 (1.80%) | 92.89 | 5M | 11.6 | -1.94 | 14 |
| **$Mn_{50}Ni_{40.5}Co_{9.5}Ti_{10}$** | ***5.916*** | **4.364 (4.33%)** | **5.437 (-8.10%)** | **4.2592 (1.82%)** | **93.24** | ***5M*** | **13.2** | **-2.54** | **This work** |
| **$Mn_{50}Ni_{42}Co_8Ti_{10}$** | ***5.915*** | **4.376 (4.62%)** | **5.420 (-8.37%)** | **4.259 (1.83%)** | **93.26** | ***5M*** | **13.7** | **-2.54** | **This work** |
| $Ni_{50}Mn_{37}Sn_{13}$ | 5.973 | 4.313 (2.11%) | 5.740 (-2.11%) | 4.2005 (-0.56%) | 90 | 4O | 4.6 | -2.40 | 5 |
| $Ni_{50}Mn_{37}Sb_{13}$ | 5.971 | 4.305 (1.96%) | 5.77 (-3.37%) | 4.2035 (-0.44%) | 90 | 4O | 4.1 | -1.90 | 5 |
| $Ni_{54}Fe_{19}Ga_{27}$ | 5.76 | 4.24 (4.10%) | 5.38 (-6.60%) | 4.1814 (2.66%) | 93.18 | 14M | 12.1 | -0.33 | 15 |
| $Ni_{50}Mn_{37}Sb_{13}$-$Ni_{50}Mn_{37.5}Sb_{12.5}$ | 5.9711 | 4.3654 (3.39%) | 5.5930 (-6.33%) | 4.2638 (1.69%) | 90 | 4O | 8.3 | -2.20 | 16 |
| $Ni_{53}Mn_{22}Ga_{25}$ | 5.811 | 4.222 (2.75%) | 5.537 (-4.72%) | 4.1982 (2.17%) | 92.5 | 14M | 9.4 | -0.06 | 17 |
| $Ni_{41}Co_{9}Mn_{40}Sn_{10}$ | 5.96 | 4.383 (4.0%) | 5.549 (-6.9%) | 4.303 (2.1%) | 90 | 6M | 9.6 | -1.13 | 18 |
| **$Mn_{50}Ni_{42}Co_{8}Ti_{10}$** | **5.915** | **5.218 (-11.78 %)** | | **7.38 (24.77%)** | | **NM** | **24.8** | **-2.88** | **This work** |
| $Mn_{2}NiGa$ | 5.9072 | 5.5272 (-6.43%) | | 6.7044 (13.49%) | | NM | 13.5 | -0.64 | 19 |
| $Ni_{40}Co_{10}Mn_{32}Al_{18}$-$Ni_{40}Co_{10}Mn_{34}Al_{16}$ | 5.824 | 5.4518 (-6.4%) | | 6.612 (13.53%) | | NM | 13.5 | -0.52 | 20 |
| $Ni_{53}Mn_{22}Ga_{25}$ | 5.811 | 5.4857 (-5.60%) | | 6.511 (12.05%) | | NM | 12.0 | -0.15 | 17 |
| $Fe_{43}Mn_{28}Ga_{29}$ | 5.864 | 5.381 (-8.24%) | | 7.058 (20.36%) | | NM | 20.4 | 1.35 | 21 |

Sym. is abbreviation of symmetry (martensite).

**TABLE S3. Lattice constants, uniaxial strains, volume changes ($\Delta V/V$) of some other typical shape memory alloys**

| **Alloys** | **Parent** | | **Martensite** | | | | | |
|---|---|---|---|---|---|---|---|---|
| | ***$a_0$*** | ***$c_0$*** | ***a* (strain)** | ***b* (strain)** | ***c* (strain)** | **\|ε\| (%)** | ***ΔV/V* (%)** | **Ref.** |
| $Fe_{3}Pt$ | 3.732 | | 3.666 (-1.77%) | | 3.777 (1.21%) | 2.94 | -2.34 | 22 |
| FePd | 3.750 (313K) | | 3.636 (193K) | | 3.860 (193K) | 5.80 | | 23 |
| $Ti_{50}Ni_{48}Fe_{2}$ | 3.0156 | | 3.0035 (-0.40%) | | 3.0395 (0.79%) | 1.18 | -0.02 | 24 |
| $Mn_{1.07}Co_{0.92}Ge$ | 4.083 | 5.285 | 5.929 (12.18%) | 3.830 (-6.20%) | 7.041 (-0.44%) | - | 4.8 | 25 |
| $Mn_{0.84}Fe_{0.16}NiGe$ | 4.0956 | 5.3546 | 6.0116 (12.27%) | 3.7438 (-8.59%) | 7.0970 (-8.47%) | - | 2.68 | 26 |
| $Mn_{0.64}Fe_{0.36}NiGe_{0.5}Si_{0.5}$ | 4.030 | 5.228 | 5.8898 (12.66%) | 3.6886 (-8.47%) | 7.0134 (0.48%) | - | 3.6 | 27 |